\documentclass[10pt,psfig]{article}
\usepackage{amssymb}
\usepackage{amsmath,amsthm}
\usepackage{amsfonts}
\usepackage{graphicx}
\usepackage{graphics}
\numberwithin{equation}{section}

\begin{document}
\title{\setlength{\baselineskip}{0.6\baselineskip}
\begin{LARGE}\textbf{Construction of quantum codes based on self-dual orientable embeddings of complete multipartite graphs}\end{LARGE}}
\author{\normalsize{Avaz Naghipour}\\
\footnotesize{Department of Computer Engineering, University
College of Nabi Akram,}\\  \footnotesize{No. 1283 Rah Ahan
Street, Tabriz, Iran}\\
\footnotesize{Department of Applied Mathematics,
Faculty of Mathematical Sciences, University of Tabriz,}\\
\footnotesize{29 Bahman Boulevard, Tabriz, Iran }\\
\footnotesize{a\_naghipour@tabrizu.ac.ir}\\
[2mm] \normalsize{
 Mohammad Ali Jafarizadeh}\\
\footnotesize{Department of Theoretical Physics and
Astrophysics, Faculty of Physics, University of Tabriz,}\\
\footnotesize{29 Bahman Boulevard, Tabriz, Iran }\\
\footnotesize{jafarizadeh@tabrizu.ac.ir}\\
[2mm] \normalsize{
 Sedaghat Shahmorad}\\
\footnotesize{Department of Applied Mathematics,
Faculty of Mathematical Sciences, University of Tabriz,}\\
\footnotesize{29 Bahman Boulevard, Tabriz, Iran }\\
\footnotesize{shahmorad@tabrizu.ac.ir}}
\date{\footnotesize{6 July 2015}}
\maketitle
\begin{small}
\hspace{-0.7cm}
\begin{quote}
\footnotesize{This paper presents four new classes of binary quantum
codes with minimum distance $3$ and $4$, namely Class-I, Class-II,
Class-III and Class-IV. The classes Class-I and Class-II are
constructed based on self-dual orientable embeddings of the complete
graphs $K_{4r+1}$ and $K_{4s}$ and by current graphs and rotation
schemes. The parameters of two classes of quantum codes are
$[[2r(4r+1),2r(4r-3),3]]$ and $[[2s(4s-1),2(s-1)(4s-1),3]]$
respectively, where $r\geq 1$ and $s\geq 2$. For these quantum
codes, the code rate approaches $1$ as $r$ and $s$ tend to infinity.
The Class-III with rate $\frac{1}{2}$ and with minimum distance $4$
is constructed by using self-dual embeddings of complete bipartite
graphs. The parameters of this class are
$[[rs,\frac{(r-2)(s-2)}{2},4]]$, where $r$ and $s$ are both
divisible by $4$. The proposed Class-IV is of minimum distance $3$
and code length $n=(2r+1)s^{2}$}. This class is constructed based on
self-dual embeddings of complete tripartite graph $K_{rs,s,s}$ and
its parameters are $[[(2r+1)s^{2},(rs-2)(s-1),3]]$, where $r\geq 2$
and $s\geq 2$.
\vspace{.3cm}
\end{quote}
\parindent 1em
\end{small}
\section{\hspace*{-0.6cm}{.} Introduction}
Quantum error-correcting codes (QECs) play an essential role in
various quantum informational processes. In the theory of quantum
computation, the information is stored in entangled states of
quantum systems. Since the interaction with the system environment
is inevitable, these interactions create noise that disrupt the
encoded data and make mistakes. One of the most useful techniques to
reduce the effects of these noise is applying the QECs. The first
quantum code $[[9,1,3]]$ was discovered by Shor [1]. Calderbank et
al. [2] introduced a systematic way for constructing the QEC from
classical error-correcting code. The problem of constructing toric
quantum codes has motivated considerable interest in the literature.
This problem was generalized within the context of surface codes [8]
and color codes [3]. The most popular toric code was proposed for
the first time by Kitaev's [5]. This code defined on a square lattic
of size $m\times m$ on the torus. The parameters of this class of
codes are $[[n,k,d]]=[[2m^2,2,m]]$. In similar way, the authors in
[7] have introduced a construction of topological quantum codes in
the projective plane $\mathbb{R}P^2$. They showed that the original
Shor's $9$-qubit repetition code is one of these codes which can be
constructed in a planar domain.
\\
\hspace*{0.5cm} Leslie proposed a new type of sparse CSS quantum
error correcting codes based on the homology of hypermaps defined on
an $m\times m$ square lattice [6]. The parameters of
hypermap-homology codes are $[[(\frac{3}{2})m^2,2,m]]$. These codes
are more efficient than Kitaev's toric codes. This seemed suggests
good quantum codes that is constructed by using hypergraphs. But
there are other surface codes with better parameters than the
$[[2m^2,2,m]]$ toric code. There exist surface codes with parameters
$[[m^2+1,2,m]]$, called homological quantum codes. These codes were
introduced by Bombin and Martin-Delgado [8].
\\
\hspace*{0.5cm} Authors in [9] presented a new class of toric
quantum codes with parameters $[[m^2,2,m]]$, where $m=2(l+1),l\geq
1$. Sarvepalli [10] studied relation between surface codes and
hypermap-homology quantum codes. He showed that a canonical hypermap
code is identical to a surface code while a noncanonical hypermap
code can be transformed to a surface code by CNOT gates alone. Li et
al. [17] introduced a large number of good binary quantum codes of
minimum distances five and six by Steane's Construction. In [18]
good binary quantum stabilizer codes obtained via graphs of Abelian
and non-Abelian groups schemes. In [19], Qian presented a new method
for constructing quantum codes from cyclic codes over finite ring
$\mathrm{F_{2}}+v\mathrm{F_{2}}$. In [20] two new classes of binary
quantum codes with minimum distance of at least three presented by
self-complementary self-dual orientable embeddings of voltage graphs
and Paley graphs. In $2013$ Korzhik [21] studied generating
nonisomorphic quadrangular embeddings of a complete graph. In [22]
M. Ellingham gave techniques to construct graph embeddings.
\\
\hspace*{0.5cm} Our aim in this work is to present four new
classes of binary quantum codes with parameters
$[[2r(4r+1),2r(4r-3),3]]$, $[[2s(4s-1),2(s-1)(4s-1),3]]$,\\
$[[rs,\frac{(r-2)(s-2)}{2},4]]$ and $[[(2r+1)s^{2},(rs-2)(s-1),3]]$
respectively, based on results of Pengelley [13], Archdeacon et al.
[14] and Archdeacon [25] in self-dual orientable embeddings of the
complete graphs $K_{4r+1}$ and $K_{4s}$, ($r\geq 1$ and $s\geq 2$),
complete bipartite graphs and complete multipartite graphs by
current graphs and rotation schemes [23]. Binary quantum codes are
defined by pair $(H_{X},H_{Z})$ of $\mathbb{Z}_{2}$-matrices with
$H_{X}H_{Z}^{T}=0$. These codes have parameters $[[n,k,d_{min}]]$,
where $k$ logical qubits are encoded into $n$ physical qubits with
minimum distance $d_{min}$. A minimum distance $d_{min}$ code can
correct all errors up to $\lfloor\frac{d_{min}-1}{2}\rfloor$ qubits.
The code rate for two classes of quantum codes of length
$n=2r(4r+1)$ and $n=2s(4s-1)$ is determined by
$\frac{k}{n}=\frac{2r(4r-3)}{2r(4r+1)}$ and
$\frac{k}{n}=\frac{2(s-1)(4s-1)}{2s(4s-1)}$, and this rate
approaches $1$ as $r$ and $s$ tend to infinity.
\\
\hspace*{0.5cm} The paper is organized as follows. The simplices
definition , chain complexes and homology group are recalled in
Section 2. In Section 3 we shall briefly present the current graphs
and rotation schemes. In Section 4, we give a brief outline of
self-dual orientable embeddings of the complete graph. Section 5 is
devoted to present new classes of binary quantum codes by using
self-dual orientable embeddings of the complete graphs $K_{4r+1}$
and $K_{4s}$, complete bipartite graphs, and complete tripartite
graph $K_{rs,s,s}$. The paper is ended with a brief conclusion.
\section{\hspace*{-0.6cm}{.} Homological algebra}
In this section, we review some fundamental notions of homology
spaces. For more detailed information about homology spaces, refer
to [4], [12].
\\
\\
\textbf{Simplices.\hspace*{1mm}} Let $m,n\in \mathbb{N}$, $m\geq n$.
Let moreover the set of points
$\{\upsilon_{0},\upsilon_{1},...,\upsilon_{n}\}$ of $\mathbb{R}^m$
be geometrically independent. An $n$-simplex $\Delta$ is a subset of
$\mathbb{R}^m$ given by
\\
\begin{equation}
\Delta=\{x\in\mathbb{R}^m|x=\sum_{i=0}^{n}t_{i}\upsilon_{i}; 0\leq
t_{i}\leq 1; \sum_{i=0}^{n}t_{i}=1\}.
\end{equation}
\\
\\
\textbf{Chain complexes.\hspace*{1mm}} Let $K$ be a simplicial
complex and $p$ a dimension. A $p$-\textit{chain} is a formal sum of
$p$-simplices in $K$. The standard notation for this is
$c=\sum_{i}n_{i}\sigma_{i}$, where $n_{i}\in \mathbb{Z}$ and
$\sigma_{i}$ is a $p$-simplex in $K$. Let $C_{p}(K)$ be the set of
all $p$-chains in $K$. The \textit{boundary homomorphism}
$\partial_{p}: C_{p}(K)\longrightarrow C_{p-1}(K)$ is defined as
\\
\begin{equation}
\partial_{k}(\sigma)=\sum_{j=0}^{k}(-1)^j[\upsilon_{0},\upsilon_{1},...,\upsilon_{j-1},\upsilon_{j+1},...,\upsilon_{k}].
\end{equation}
\\
The \textit{chain complex} is the sequence of chain groups connected
by boundary homomorphisms,
\\
\begin{equation}
\cdots
\stackrel{\partial_{p+2}}{\longrightarrow}C_{p+1}\stackrel{\partial_{p+1}}{\longrightarrow}C_{p}
\stackrel{\partial_{p}}{\longrightarrow}C_{p-1}\stackrel{\partial_{p-1}}{\longrightarrow}\cdots
\end{equation}
\\
\textbf{Cycles and boundaries.\hspace*{1mm}} We are interested in
two subgroups of $C_{p}(K)$, \textit{cycle} and \textit{boundary}
groups. The $p$-th cycle group is the kernel of
$\partial_{p}:C_{p}(K)\longrightarrow C_{p-1}(K)$, and denoted as
$Z_{p}=Z_{p}(K)$. The $p$-th boundary group is the image of
$\partial_{p+1}:C_{p+1}(K)\longrightarrow C_{p}(K)$, and denoted as
$B_{p}=B_{p}(K)$.
\\
\\
\textbf{Definition 2.1\hspace*{1mm}}(Homology group, Betti number).
The $p$-\textit{th homology group} $H_{p}$ is the $p$-th cycle group
modulo the $p$-th boundary group, $H_{p}=Z_{p}/B_{p}$. The
$p$-\textit{th Betti number} is the rank (i.e. the number of
generators) of this group, $\beta_{p}$=rank $H_{p}$. So the first
homology group $H_{1}$ is given as
\\
\begin{equation}
H_{1}=Z_{1}/B_{1}.
\end{equation}
\\
From the algebraic topology, we can see that the group $H_{1}$ only
depends, up to isomorphisms, on the topology of the surface [4]. In
fact
\\
\begin{equation}
H_{1}\simeq \mathbb{Z}_{2}^{2g},
\end{equation}
\\
where $g$ is the genus of the surface, i.e. the number of ``holes''
or ``handles''. We then have
\\
\begin{equation}
|H_{1}|=2^{2g}.
\end{equation}
\section{\hspace*{-0.6cm}{.} Current graphs}
Current graphs were invented before voltage graphs. These graphs
are used in proof of Map Color Theorem, determination of minimum
genus of complete graphs. Current graphs are dual of voltage
graphs which apply to \textit{embedded} voltage graphs. Faces of
current graph correspond to vertices in voltage graph, and vice
versa.  We refer the reader to [23] for more details.
\subsection{\hspace*{-0.5cm}{.} Rotation schemes}
Rotation schemes are important for applying the face-tracing. Let
$G=(V,E)$ be a connected graph. Denote the vertex set of $G$ by
$V(G)=\{1,2,...,n\}$. For each $i\in V(G)$, let $V(i)=\{k\in
V(G)|\{i,k\}\in E(G)\}$. Let $p_{i}:V(i)\rightarrow V(i)$ be an
oriented cyclic order (or a cyclic permutation) on $V(i)$, of
length $n_{i}=|V(i)|$; $p_{i}$ ia called a \textit{rotation
scheme}, or \textit{rotation system}. By combining the two
concepts of rotation schemes and $2$-cell embeddings we have the
following theorem:
\\
\\
\textbf{Theorem 3.1.\hspace*{1mm}}Every rotation scheme for a
graph $G$ induces a unique embedding of $G$ into an orientable
surface. Conversely, every embedding of a graph into orientable
surface induces a unique rotation scheme for $G$.
\\
\\
\textbf{Proof.\hspace*{1mm}} The Proof of this theorem is found in
[16].
\\
\\
In proof of this theorem $D^{*}=\{(a,b)|\{a,b\}\in E(G)\}$, and
$P^{*}$ as a permutation on the set $D^{*}$ of directed edges of $G$
is defined as
\\
\begin{equation}
P^{*}(a,b)=(b,p_{b}(a)).
\end{equation}
\\
The orbits under $P^{*}$ determine the ($2$-cell)faces of the
corresponding embedding. \\
\hspace*{0.5cm} Often it is customary to represent a graph $G$
with rotation in the plane in such a way that a clockwise (or
counterclockwise) reading of the edges incident with a vertex
gives the rotation at that vertex. By convention, a solid vertex
has its incident edges ordered clockwise; a hollow vertex,
counterclockwise. For further information about rotation schemes,
the reader is referred to Refs. [16, 24].
\section{\hspace*{-0.6cm}{.} Self-dual orientable embeddings of
 complete graph}
Let $S$ be a compact, connected, oriented surface (i.e.
$2$-manifold) with genus $g$. As shown by Pengelley [13], Euler's
Formula excludes self-dual orientable embeddings of $K_{n}$ unless
$n\equiv0$ or $1$ (mod $4$). Let $T$ be a self-dual embedding of
$S$ and $\alpha_{0}$, $\alpha_{1}$, $\alpha_{2}$ denote
respectively the number of vertices, edges and faces $T$. If the
embedding is in $S$, then
\\
\begin{equation}
\alpha_{2}=\alpha_{0}=n,
\hspace*{1.5cm}\alpha_{1}=\frac{n(n-1)}{2},
\end{equation}
\\
and Euler's Formula follows that
\\
\begin{equation}
g=\frac{2-\alpha_{0}+\alpha_{1}-\alpha_{2}}{2}=\frac{(n-1)(n-4)}{4}.
\end{equation}
\\
Hence such an embedding can exist only if $n\equiv0$ or $1$ (mod
$4$). Since the embedding consists of $n$ faces, each face must be
adjacent to every other along exactly one edge. Hence each face is
an $(n-1)$-gon. Pengelley presented how to use current graphs and
rotation schemes to describe an orientable embedding of $K_{n}$
having $n$ $(n-1)$-gons as faces, and then show such an embedding is
self-dual [13]. Since we wish our faces to be $(n-1)$-gons, by the
current graph construction principles is requiring that each vertex
in the current graph be of valence $n-1$ [27].\\
In the case $n\equiv1$ (mod $4$), we select a group $\Gamma$ for
which $K_{4r+1}$ is a Cayley color graph; in this case, we can only
pick $\Gamma=\mathbb{Z}_{4r+1}$. Label the vertices $K_{4r+1}$ with
the elements of $\Gamma$. Denote the faces by the numbers $0$,
$1$,$\ldots$,$n-1$. Then choose a certain orientation for each face.
Write down the cyclic order of the faces adjacent to face $0$. This
gives a certain permutation of $1$, $2$,$\ldots$,$n-1$. Do the same
for the other faces. This leads to a scheme. The scheme for
$K_{4r+1}$ is generated by the following $0$ [13].
\\
\begin{equation*}
0.\hspace*{1mm} 1, -2, -1, 2, 3, -4, -3, 4,\ldots,2k-1, -2k,
-(2k-1), 2k,\ldots,2r-1, -2r, -(2r-1), 2r\hspace*{2mm}
\end{equation*}
\begin{equation}
\hspace*{8cm} (1\leq k\leq r).
\end{equation}
\\
Explicit calculation gives the following cyclic sequence for the
vertices of the face containing the directed edge from $0$ to $1$:
\\
\begin{equation*}
0, 1, 3, 2, 0,\ldots,0, 2k-1, 4k-1, 2k, 0,\ldots,0, 2r-1, 4r-1,
2r\hspace*{2mm}
\end{equation*}
\begin{equation}
\hspace*{8cm} (1\leq k\leq r).
\end{equation}
\\
In general, by explicit calculation from scheme, we obtain, for
each $i\in \mathbb{Z}_{4r+1}$, the cyclic sequence
\begin{equation*}
i, 1+i, 3+i, 2+i, i,\ldots,i, (2k-1)+i, (4k-1)+i, 2k+i,
i,\ldots,i, (2r-1)+i, (4r-1)+i, 2r+i\hspace*{2mm}
\end{equation*}
\begin{equation}
\hspace*{8cm} (1\leq k\leq r).
\end{equation}
\\
for the vertices of the face containing the directed edge $e_{i}$,
where $e_{i}$ is an edge between vertices $i$ and $i+1$.\\
In the case $n\equiv0$ (mod $4$), Pengelly [13] has used the
following group
\\
\begin{equation*}
G=\underbrace{\mathbb{Z}_{2}\times \mathbb{Z}_{2}\times
\cdots\times\mathbb{Z}_{2}}_{s\hspace*{1mm}times}\times
\mathbb{Z}_{t}
\end{equation*}
\\
where $n=2^st$, $t$ is odd, and $s\geq 2$. There are precisely
$2^s-1$ elements of order two. Pengelly in Ref. [13] released
$2^{s}(t-1)$ elements with distinct inverses, and his choose
$a_{1}$, $a_{2}$, $a_{3}$,$\ldots$,$a_{2^{s-1}(t-1)}$ from this
collection such that they and their inverses deplete the collection.
We label the elements of order two $b_{1}$, $b_{2}$,
$b_{3}$,$\ldots$,$b_{2^s-1}$. From the elementary group theory, we
can see that $\sum_{l=1}^{2^{s}-1}b_{l}=0$. The scheme for
$K_{4s}(s\geq 2)$ is generated by the following $0$ [13].
\\
\begin{equation*}
0.\hspace*{1mm} a_{1}, -a_{2}, -a_{1}, a_{2}, a_{3}, -a_{4},
-a_{3}, a_{4},\ldots,a_{2k-1}, -a_{2k}, -a_{(2k-1)},
a_{2k},\ldots,a_{2^{s-1}(t-1)-1},
\end{equation*}
\begin{equation*}
\hspace*{-3.6cm} -a_{2^{s-1}(t-1)}, -a_{2^{s-1}(t-1)-1},
a_{2^{s-1}(t-1)},b_{1}, b_{2},\ldots,b_{2^{s}-1}\hspace*{2mm}
\end{equation*}
\begin{equation}
\hspace*{8cm} (1\leq k\leq 2^{s-2}(t-1)).
\end{equation}
\\
This generates by the additive rule the entire scheme for
$K_{4s}$. Explicit calculation yield the following cyclic sequence
for the vertices of the face containing the directed edge from $0$
to $a_{1}$ [13]:
\\
\begin{equation*}
0, a_{1}, a_{1}+a_{2}, a_{2}, 0, a_{3}, a_{3}+a_{4}, a_{4},
0,\ldots,0, a_{2k-1}, a_{2k-1}+a_{2k}, a_{2k}, 0,\ldots,0,
a_{2^{s-1}(t-1)-1}
\end{equation*}
\begin{equation*}
\hspace*{-3.4cm} +a_{2^{s-1}(t-1)}, a_{2^{s-1}(t-1)}, 0, b_{1},
b_{1}+b_{2},
b_{1}+b_{2}+b_{3},\ldots,\sum_{l=1}^{2^{s}-2}b_{l}\hspace*{2mm}
\end{equation*}
\begin{equation}
\hspace*{8cm} (1\leq k\leq 2^{s-2}(t-1)).
\end{equation}
\\
By calculation from scheme, we obtain, for each $g\in G$, the
cyclic sequence for the vertices of the face containing the
directed edge from $g$ to $g+a_{1}$:
\\
\begin{equation*}
g, a_{1}+g, a_{1}+a_{2}+g, a_{2}+g, g, a_{3}+g, a_{3}+a_{4}+g,
a_{4}+g, g,\ldots,g, a_{2k-1}+g, a_{2k-1}+a_{2k}+g,
\end{equation*}
\begin{equation*}
\hspace*{-1.4cm} g,\ldots,g,
a_{2^{s-1}(t-1)-1}+a_{2^{s-1}(t-1)}+g, a_{2^{s-1}(t-1)}+g,
g,b_{1}+g, b_{1}+b_{2}+g,
\end{equation*}
\begin{equation}
\hspace*{-4cm}b_{1}+b_{2}+b_{3}+g,\ldots,g+\sum_{l=1}^{2^{s}-2}b_{l}\hspace*{4mm}(1\leq
k\leq 2^{s-2}(t-1)).
\end{equation}
\\
\section{\hspace*{-0.6cm}{.} Quantum codes from graphs on surfaces}
The idea of constructing CSS (Calderbank-Shor-Steane) codes from
graphs embedded on surfaces has been discussed in a number of
papers. See for detailed descriptions e.g. [11]. Let $X$ be a
compact, connected, oriented surface (i.e. 2-manifold) with genus
$g$. A tiling of $X$ is defined to be a cellular embedding of an
undirected (simple) graph $G=(V,E)$ in a surface. This embedding
defines a set of faces $F$. Each face is described by the set of
edges on its boundary. This tiling of surface is denoted
$M=(V,E,F)$. The dual graph $G$ is the graph $G^*=(V^*,E^*)$ such
that:
\\
\\
i) One vertex of $G^*$ inside each face of $G$,
\\
\\
ii) For each edge $e$ of $G$ there is an edge $e^*$ of $G^*$ between
the two vertices of \hspace*{4mm} $G^*$ corresponding to the two
faces of $G$ adjacent to $e$.
\\
\\
It can be easily seen that, there is a bijection between the edges
of $G$ and the edges of $G^*$.
\\
\\
\hspace*{0.5cm} There is an interesting relationship between the
number of elements of a lattice embedded in a surface and its genus.
The Euler characteristic of $\chi$ is defined as its number of
vertices ($|V|$) minus its number of edges ($|E|$) plus its number
of faces ($|F|$), i.e.,
\\
\begin{equation}
\chi=|V|-|E|+|F|.
\end{equation}
\\
For closed orientable surfaces we have
\\
\begin{equation}
\chi=2(1-g).
\end{equation}
\\
The \textit{surface code} associated with a tiling $M=(V,E,F)$ is
the CSS code defined by the matrices $H_{X}$ and $H_{Z}$ such that
$H_{X}\in \mathcal{M}_{|V|,|E|}(\mathbb{Z}_{2})$ is the vertex-edge
incidence matrix of the tiling and $H_{Z}\in
\mathcal{M}_{|F|,|E|}(\mathbb{Z}_{2})$ is the face-edge incidence
matrix of the tiling. Therefore, from $(X,G)$ is constructed a CSS
code with parameters $[[n,k,d]]$. where $n$ is the number of edges
of $G$, $k=2g$ (by (2.6)) and $d$ is the shortest non-boundary cycle
in $G$ or $G^*$. In this work, the minimum distance of quantum codes
by a parity check matrix $H$ (or generator matrix) is obtained. For
a detailed information to compute the minimum distance, we refer the
reader to [15].
\subsection{\hspace*{-0.5cm}{.} New class of $[[2r(4r+1),2r(4r-3),3]]$ binary quantum codes \hspace*{-0.3cm} from
embeddings of $K_{4r+1}$} Our aim in this subsection is to construct
a new class of binary quantum codes by using self-dual orientable
embeddings of complete graphs. Let $G=K_{m}$ be an embedding in an
orientable surface $S$. We know that $|E(G)|=\frac{m(m-1)}{2}$.
Also, from (4.2) with a self-dual embedding of complete graph on an
orientable surface of genus $g$, we know that if $m=4r+1\equiv1$
(mod $4$), then $g=\frac{(m-1)(m-4)}{4}=r(4r-3)$. Therefore,
$|E(G)|=2r(4r+1)$. By finding the vertex-edge incidence matrix
$H_{X}$ using the relation (4.3) and rotation schemes, and the
face-edge incidence matrix $H_{Z}$ by using (4.5), one can easily
see that $H_{X}H_{Z}^{T}=0$ and $d_{min}=3$. Thus the code
parameters are given by: the code minimum distance is $d_{min}=3$;
the code length is $n=|E(G)|=2r(4r+1)$ and $k=2g=2r(4r-3)$.
Consequently, the class of codes with parameters
$[[2r(4r+1),2r(4r-3),3]]$,$(r\geq 1)$ is obtained.
\\
\\
\textbf{Example 5.1.1.\hspace*{1mm}} Let $m=5=4\times 1+1\equiv1$
(mod $4$). Then $n=|E(G)|=10$ and $k=2g=2$. For determining
$d_{min}$ by using rotation schemes, by the following Figure $1$
we have:
\\
\\
$$
\setlength{\unitlength}{1cm}
\begin{picture}(3,3)
\linethickness{0.095mm}\put(1.5,-0.3){$a$}
 \put(3.2,1.5){$b$}
 \put(-0.3,1.5){$b$}
 \put(1.5,3.2){$a$}
 \put(-0.4,-0.3){$0$}\put(0,0){\circle*{0.2}}
 \put(-0.3,3.1){$0$}\put(0,3){\circle*{0.2}}
 \put(3.2,-0.3){$0$}\put(3,0){\circle*{0.2}}
 \put(3.1,3.1){$0$}\put(3,3){\circle*{0.2}}
\put(0,3){\line(0,-3){1.5}} \put(0,0){\line(0,3){1.5}}
\put(3,3){\line(0,-3){1.5}} \put(3,0){\line(0,4){1.5}}
\put(0,0){\line(6,0){3}}\put(0,3){\line(6,0){3}}
\put(0,3){\line(1,-2){1.5}} \put(0,0){\line(2,1){3}}
\put(3,3){\line(-2,-1){3}}\put(3,0){\line(-1,2){1.5}}
\put(1.2,0.9){$3$}\put(1.2,0.6){\circle*{0.2}}
\put(2.4,1.4){$1$}\put(2.4,1.2){\circle*{0.2}}
\put(0.6,2.1){$4$}\put(0.6,1.8){\circle*{0.2}}
\put(1.8,2.7){$2$}\put(1.8,2.4){\circle*{0.2}}
\end{picture}
$$
\\[-0.5cm]
\begin{figure}[htbp]
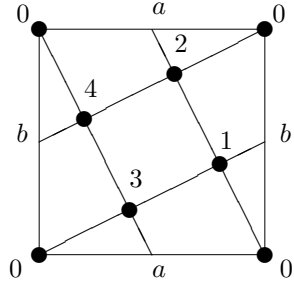

\vspace{-0.7cm} \caption{\small{A rotation embedding $K_{5}$ in
the torus}} \label{table:2}
\end{figure}
\\
\\
\begin{equation*}
0.\hspace*{2mm}2\hspace*{2mm}4\hspace*{2mm}3\hspace*{2mm}1
\end{equation*}
\begin{equation*}
1.\hspace*{2mm}3\hspace*{2mm}0\hspace*{2mm}4\hspace*{2mm}2
\end{equation*}
\begin{equation*}
2.\hspace*{2mm}4\hspace*{2mm}1\hspace*{2mm}0\hspace*{2mm}3
\end{equation*}
\begin{equation*}
3.\hspace*{2mm}0\hspace*{2mm}2\hspace*{2mm}1\hspace*{2mm}4
\end{equation*}
\begin{equation*}
4.\hspace*{2mm}1\hspace*{2mm}3\hspace*{2mm}2\hspace*{2mm}0
\end{equation*}
\\
\\
By rotation schemes, we get the following vertex-edge and
face-edge matrices respectively:
\\
\\
\begin{equation*}
H_{X}=\left(
\begin{array}{cccccccccc}
1 & 1 & 1 & 1 & 0 & 0 & 0 & 0 & 0 & 0\\
1 & 0 & 0 & 0 & 1 & 1 & 1 & 0 & 0 & 0\\
0 & 1 & 0 & 0 & 1 & 0 & 0 & 1 & 1 & 0\\
0 & 0 & 1 & 0 & 0 & 1 & 0 & 1 & 0 & 1\\
0 & 0 & 0 & 1 & 0 & 0 & 1 & 0 & 1 & 1
\end{array}\right)
\end{equation*}
\\
\begin{equation*}
H_{Z}=\left(
\begin{array}{cccccccccc}
1 & 1 & 0 & 0 & 0 & 1 & 0 & 1 & 0 & 0\\
1 & 0 & 1 & 0 & 0 & 0 & 1 & 0 & 0 & 1\\
0 & 0 & 1 & 1 & 0 & 0 & 0 & 1 & 1 & 0\\
0 & 1 & 0 & 1 & 1 & 0 & 1 & 0 & 0 & 0\\
0 & 0 & 0 & 0 & 1 & 1 & 0 & 0 & 1 & 1
\end{array}\right)
\end{equation*}
\\
\\
One can easily see that $H_{X}H_{Z}^{T}=0$ and $d_{min}=3$.
Therefore, the code with parameters $[[10,2,3]]$ is obtained.
\\
In the above figure the torus is recovered from the rectangle by
identifying its left and right sides and simultaneously
identifying its top and bottom sides.
\\
\\
\textbf{Example 5.1.2.\hspace*{1mm}} Let $m=9=4\times 2+1\equiv1$
(mod $4$). Then $n=|E(G)|=36$ and $k=2g=20$. For determining
$H_{X}$ and $H_{Z}$ by using (4.3) and rotation schemes, we have:
\\
\\
\begin{equation*}
0.\hspace*{2mm}1\hspace*{2mm}7\hspace*{2mm}8\hspace*{2mm}2\hspace*{2mm}3\hspace*{2mm}5
\hspace*{2mm}6\hspace*{2mm}4
\end{equation*}
\begin{equation*}
1.\hspace*{2mm}2\hspace*{2mm}8\hspace*{2mm}0\hspace*{2mm}3\hspace*{2mm}4\hspace*{2mm}6
\hspace*{2mm}7\hspace*{2mm}5
\end{equation*}
\begin{equation*}
2.\hspace*{2mm}3\hspace*{2mm}0\hspace*{2mm}1\hspace*{2mm}4\hspace*{2mm}5\hspace*{2mm}7
\hspace*{2mm}8\hspace*{2mm}6
\end{equation*}
\begin{equation*}
3.\hspace*{2mm}4\hspace*{2mm}1\hspace*{2mm}2\hspace*{2mm}5\hspace*{2mm}6\hspace*{2mm}8
\hspace*{2mm}0\hspace*{2mm}7
\end{equation*}
\begin{equation}
4.\hspace*{2mm}5\hspace*{2mm}2\hspace*{2mm}3\hspace*{2mm}6\hspace*{2mm}7\hspace*{2mm}0
\hspace*{2mm}1\hspace*{2mm}8
\end{equation}
\begin{equation*}
5.\hspace*{2mm}6\hspace*{2mm}3\hspace*{2mm}4\hspace*{2mm}7\hspace*{2mm}8\hspace*{2mm}1
\hspace*{2mm}2\hspace*{2mm}0
\end{equation*}
\begin{equation*}
6.\hspace*{2mm}7\hspace*{2mm}4\hspace*{2mm}5\hspace*{2mm}8\hspace*{2mm}0\hspace*{2mm}2
\hspace*{2mm}3\hspace*{2mm}1
\end{equation*}
\begin{equation*}
7.\hspace*{2mm}8\hspace*{2mm}5\hspace*{2mm}6\hspace*{2mm}0\hspace*{2mm}1\hspace*{2mm}3
\hspace*{2mm}4\hspace*{2mm}2
\end{equation*}
\begin{equation*}
8.\hspace*{2mm}0\hspace*{2mm}6\hspace*{2mm}7\hspace*{2mm}1\hspace*{2mm}2\hspace*{2mm}4
\hspace*{2mm}5\hspace*{2mm}3
\end{equation*}
\\
\\
Also, by using (3.1), (4.4) and (4.5), we obtain the following
cyclic sequence for the vertices of the face containing the directed
edge for each $i\in \mathbb{Z}_{9}$
\\
\\
\begin{equation*}
(0)\hspace*{2mm}0\hspace*{2mm}1\hspace*{2mm}3\hspace*{2mm}2\hspace*{2mm}0\hspace*{2mm}3
\hspace*{2mm}7\hspace*{2mm}4
\end{equation*}
\begin{equation*}
(1)\hspace*{2mm}1\hspace*{2mm}2\hspace*{2mm}4\hspace*{2mm}3\hspace*{2mm}1\hspace*{2mm}4
\hspace*{2mm}8\hspace*{2mm}5
\end{equation*}
\begin{equation*}
(2)\hspace*{2mm}2\hspace*{2mm}3\hspace*{2mm}5\hspace*{2mm}4\hspace*{2mm}2\hspace*{2mm}5
\hspace*{2mm}0\hspace*{2mm}6
\end{equation*}
\begin{equation*}
(3)\hspace*{2mm}3\hspace*{2mm}4\hspace*{2mm}6\hspace*{2mm}5\hspace*{2mm}3\hspace*{2mm}6
\hspace*{2mm}1\hspace*{2mm}7
\end{equation*}
\begin{equation}
(4)\hspace*{2mm}4\hspace*{2mm}5\hspace*{2mm}7\hspace*{2mm}6\hspace*{2mm}4\hspace*{2mm}7
\hspace*{2mm}2\hspace*{2mm}8
\end{equation}
\begin{equation*}
(5)\hspace*{2mm}5\hspace*{2mm}6\hspace*{2mm}8\hspace*{2mm}7\hspace*{2mm}5\hspace*{2mm}8
\hspace*{2mm}3\hspace*{2mm}0
\end{equation*}
\begin{equation*}
(6)\hspace*{2mm}6\hspace*{2mm}7\hspace*{2mm}0\hspace*{2mm}8\hspace*{2mm}6\hspace*{2mm}0
\hspace*{2mm}4\hspace*{2mm}1
\end{equation*}
\begin{equation*}
(7)\hspace*{2mm}7\hspace*{2mm}8\hspace*{2mm}1\hspace*{2mm}0\hspace*{2mm}7\hspace*{2mm}1
\hspace*{2mm}5\hspace*{2mm}2
\end{equation*}
\begin{equation*}
(8)\hspace*{2mm}8\hspace*{2mm}0\hspace*{2mm}2\hspace*{2mm}1\hspace*{2mm}8\hspace*{2mm}2
\hspace*{2mm}6\hspace*{2mm}3
\end{equation*}
\\
By using (5.3) and (5.4) we obtain the following vertex-edge and
face-edge matrices respectively:
\\
\begin{equation*}
H_{X}=(A_1 \vert A_2)
\end{equation*}
\\
where
\\
\begin{equation*}
A_{1}=\left(
\begin{array}{ccccccccccccccccccc}
1 & 1 & 1 & 1 & 1 & 1 & 1 & 1 & 0 & 0 & 0 & 0 & 0 & 0 & 0 & 0 & 0 & 0\\
1 & 0 & 0 & 0 & 0 & 0 & 0 & 0 & 1 & 1 & 1 & 1 & 1 & 1 & 1 & 0 & 0
& 0\\
0 & 1 & 0 & 0 & 0 & 0 & 0 & 0 & 1 & 0 & 0 & 0 & 0 & 0 & 0 & 1 & 1 &
1\\
0 & 0 & 1 & 0 & 0 & 0 & 0 & 0 & 0 & 1 & 0 & 0 & 0 & 0 & 0 & 1 & 0 &
0\\
0 & 0 & 0 & 1 & 0 & 0 & 0 & 0 & 0 & 0 & 1 & 0 & 0 & 0 & 0 & 0 & 1 &
0\\
0 & 0 & 0 & 0 & 1 & 0 & 0 & 0 & 0 & 0 & 0 & 1 & 0 & 0 & 0 & 0 & 0 &
1\\
0 & 0 & 0 & 0 & 0 & 1 & 0 & 0 & 0 & 0 & 0 & 0 & 1 & 0 & 0 & 0 & 0 &
0\\
0 & 0 & 0 & 0 & 0 & 0 & 1 & 0 & 0 & 0 & 0 & 0 & 0 & 1 & 0 & 0 & 0 &
0\\
0 & 0 & 0 & 0 & 0 & 0 & 0 & 1 & 0 & 0 & 0 & 0 & 0 & 0 & 1 & 0 & 0 &
0
\end{array}\right)
\end{equation*}
\\
\\
\begin{equation*}
A_{2}=\left(
\begin{array}{ccccccccccccccccccc}
0 & 0 & 0 & 0 & 0 & 0 & 0 & 0 & 0 & 0
& 0 & 0 & 0 & 0 & 0 & 0 & 0 & 0\\
0 & 0 & 0 & 0 & 0 & 0 & 0 & 0 & 0 & 0 & 0 & 0 & 0 & 0 & 0 & 0 & 0
& 0\\
1 & 1 & 1 & 0 & 0 & 0 & 0 & 0 & 0 & 0 & 0 & 0 & 0 & 0 & 0 & 0 & 0 &
0\\
0 & 0 & 0 & 1 & 1 & 1 & 1 & 1 & 0 & 0 & 0 & 0 & 0 & 0 & 0 & 0 & 0 &
0\\
0 & 0 & 0 & 1 & 0 & 0 & 0 & 0 & 1 & 1 & 1 & 1 & 0 & 0 & 0 & 0 & 0 &
0\\
0 & 0 & 0 & 0 & 1 & 0 & 0 & 0 & 1 & 0 & 0 & 0 & 1 & 1 & 1 & 0 & 0 &
0\\
1 & 0 & 0 & 0 & 0 & 1 & 0 & 0 & 0 & 1 & 0 & 0 & 1 & 0 & 0 & 1 & 1 &
0\\
0 & 1 & 0 & 0 & 0 & 0 & 1 & 0 & 0 & 0 & 1 & 0 & 0 & 1 & 0 & 1 & 0 &
1\\
0 & 0 & 1 & 0 & 0 & 0 & 0 & 1 & 0 & 0 & 0 & 1 & 0 & 0 & 1 & 0  &1 &
1
\end{array}\right)
\end{equation*}
\\
\begin{equation*}
H_{Z}=(B_1 \vert B_2)
\end{equation*}
\\
where
\\
\begin{equation*}
B_{1}=\left(
\begin{array}{ccccccccccccccccccc}
1 & 1 & 1 & 1 & 0 & 0 & 0 & 0 & 0 & 1 & 0 & 0 & 0 & 0 & 0 & 1 & 0 & 0 \\
0 & 0 & 0 & 0 & 0 & 0 & 0 & 0 & 1 & 1 & 1 & 1 & 0 & 0  &0 & 0 & 1 & 0 \\
0 & 0 & 0 & 0 & 1 & 1 & 0 & 0 & 0 & 0 & 0 & 0 & 0 & 0 & 0 & 1 & 1 &
1\\
0 & 0 & 0 & 0 & 0 & 0 & 0 & 0 & 0 & 0 & 0 & 0 & 1 & 1 & 0 & 0 & 0 &
0\\
0 & 0 & 0 & 0 & 0 & 0 & 0 & 0 & 0 & 0 & 0 & 0 & 0 & 0 & 0 & 0 & 0 &
0\\
0 & 0 & 1 & 0 & 1 & 0 & 0 & 0 & 0 & 0 & 0 & 0 & 0 & 0 & 0 & 0 & 0 &
0\\
0 & 0 & 0 & 1 & 0 & 1 & 1 & 1 & 0 & 0 & 1 & 0 & 1 & 0 & 0 & 0 & 0 &
0\\
1 & 0 & 0 & 0 & 0 & 0 & 1 & 0 & 0 & 0 & 0 & 1 & 0 & 1 & 1 & 0 & 0 &
1\\
0 & 1 & 0 & 0 & 0 & 0 & 0 & 1 & 1 & 0 & 0 & 0 & 0 & 0 & 1 & 0 & 0 &
0
\end{array}\right)
\end{equation*}
\\
\\
\begin{equation*}
B_{2}=\left(
\begin{array}{ccccccccccccccccccc}
0 & 0 & 0 & 0 & 0 & 0 & 1 & 0 & 0 & 0 & 1 & 0 & 0 & 0 & 0 & 0 & 0 & 0\\
0 & 0 & 0 & 1 & 0 & 0 & 0 & 0 & 0 & 0 & 0 & 1 & 0 & 0 & 1 & 0 & 0 & 0\\
1 & 0 & 0 & 0 & 1 & 0 & 0 & 0 & 1 & 0 & 0 & 0 & 0 & 0 & 0 & 0 & 0 &
0\\
0 & 0 & 0 & 1 & 1 & 1 & 1 & 0 & 0 & 1 & 0 & 0 & 1 & 0 & 0 & 0 & 0 &
0\\
0 & 1 & 1 & 0 & 0 & 0 & 0 & 0 & 1 & 1 & 1 & 1 & 0 & 1 & 0 & 1 & 0 &
0\\
0 & 0 & 0 & 0 & 0 & 0 & 0 & 1 & 0 & 0 & 0 & 0 & 1 & 1 & 1 & 0 & 1 &
1\\
0 & 0 & 0 & 0 & 0 & 0 & 0 & 0 & 0 & 0 & 0 & 0 & 0 & 0 & 0 & 1 & 1 &
0\\
0 & 1 & 0 & 0 & 0 & 0 & 0 & 0 & 0 & 0 & 0 & 0 & 0 & 0 & 0 & 0 & 0 &
1\\
1 & 0 & 1 & 0 & 0 & 1 & 0 & 1 & 0 & 0 & 0 & 0 & 0 & 0 & 0 & 0 & 0 &
0
\end{array}\right)
\end{equation*}
\\
One can easily see that $H_{X}H_{Z}^{T}=0$ and $d_{min}=3$.
Therefore, the code with parameters $[[36,20,3]]$ is obtained.
\subsection{\hspace*{-0.5cm}{.} New class of $[[2s(4s-1),2(s-1)(4s-1),3]]$ binary quantum \hspace*{-0.3cm} codes from
embeddings of $K_{4s}$}The construction of this class will be
based on self-dual orientable embeddings of complete graph
$K_{4s}$.
\\
\\
Let $G=(V,E)$ be a self-dual graph on $4s$ vertices. From (4.1),
we know that $|E(G)|=2s(4s-1)$. Also, from (4.2) we know that
$g=(s-1)(4s-1)$. Therefore, the code length is $n=|E(G)|=2s(4s-1)$
and $k=2g=2(s-1)(4s-1)$. With finding the matrices $H_{X}$ and
$H_{Z}$ using the relations (4.6), (4.7) and (4.8), one can see
that $H_{X}H_{Z}^{T}=0$ and $d_{min}=3$. Consequently, the class
of codes with parameters $[[2s(4s-1),2(s-1)(4s-1),3]]$, ($s\geq
2$) is constructed.
\subsection{\hspace*{-0.5cm}{.} New class of $[[rs,\frac{(r-2)(s-2)}{2},4]]$ binary quantum codes \hspace*{-0.3cm} from
embeddings of complete bipartite graphs} Let the complete
bipartite graph $G=K_{r,s}$ be an embedding in $S_{g}$(the
orientable $2$-manifold of genus $g$). The graph $K_{r,s}$ has
$r+s$ vertices divided into two subsets, one of size $r$ and the
other of size $s$. The number of edges in a complete bipartite
graph $K_{r,s}$ is $|E|=rs$. From Ref. [23] and the following
theorem, the genus of the complete bipartite graph $K_{r,s}$ is
given as
\\
\begin{equation}
g=\frac{(r-2)(s-2)}{4}
\end{equation}
\\
where $r$ and $s$ are both divisible by $4$
\\
\\
\textbf{Theorem 5.3.1.\hspace*{1mm}}$K_{r,s}$ has both an
orientable and a nonorientable self-dual embedding for all even
integers $r$ and $s$ exceeding $2$, except that there is no
orientable self-dual embedding of $K_{6,6}$.
\\
\\
\textbf{Proof.\hspace*{1mm}} The Proof of this theorem is found in
[14].
\\
\\
In the complete bipartite graph $G=K_{r,s}$, where $r$ and $s$ are
both divisible by $4$, we know that $|E(G)|=rs$. Since in this
self-dual embedding on an orientable surface the code minimum
distance is four. Thus, the code parameters are given by: the code
minimum distance $d_{min}=4$; the code length is $n=|E(G)|=rs$ and
$k=2g=\frac{(r-2)(s-2)}{2}$. Consequently, the new class of codes
with parameters $[[rs,\frac{(r-2)(s-2)}{2},4]]$ is obtained.
\\
\\
\newpage
\hspace*{-0.5cm}\textbf{Example 5.3.1.\hspace*{1mm}} In Figure $2$
we give a self-dual embedding of $K_{4,4}$ into the torus. In this
figure the top of the rectangle is identified with the bottom and
the left with the right to recover the torus.
\\[-0.8cm]
\begin{figure}[htb]
\centering
\includegraphics[width=10cm]{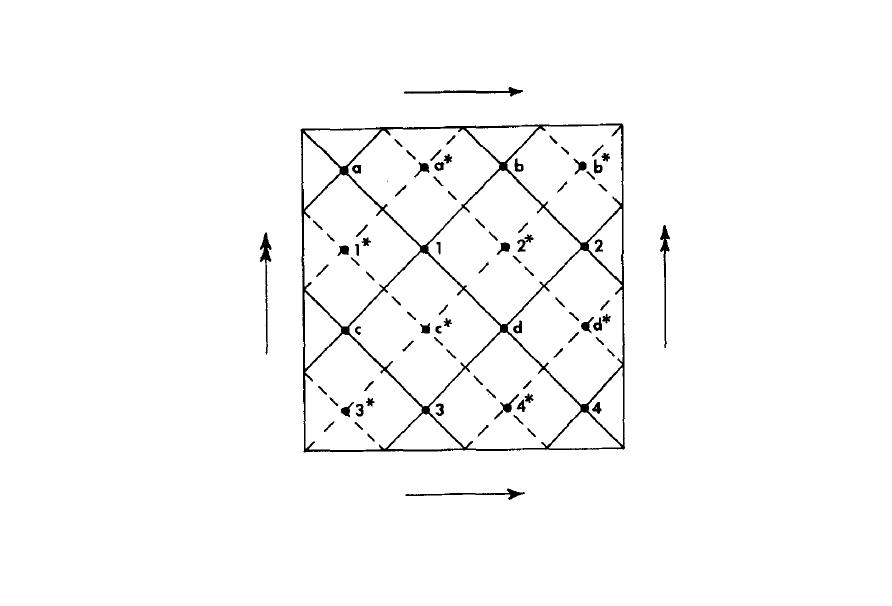}\\
\end{figure}
\\[-1.75cm]
\begin{figure}[htbp]
\vspace{-0.7cm} \caption{\small{A self-dual embedding of $K_{4,4}$
in the torus}} \label{table:2}
\end{figure}
\\
In this figure the primal graph is shown by solid lines, while the
dual graph is drawn with dotted lines. The vertex bipartitions are
$\{a, b, c, d\}$ vs. $\{1, 2, 3, 4\}$. Since in this figure the
shortest non-boundary cycle in the primal graph or the dual graph
is $4$. So, $d_{min}=4$. Also, we have $n=4\times 4=16$, the
number of edges and $k=2g=2\times \frac{(4-2)(4-2)}{4}$ (by
(5.3)). Hence, a $[[16,2,4]]$ code is obtained.
\subsection{\hspace*{-0.49cm}{.} New class of $[[(2r+1)s^{2},(rs-2)(s-1),3]]$ binary \hspace*{-0.1cm} quantum codes from
embeddings of complete tripartite graph $K_{rs,s,s}$} Let
$G=K_{rs,s,s}$ be a complete tripartite embedded on a closed
surface. The graph $K_{rs,s,s}$ has $rs+s+s$ vertices.  The number
of edges in a complete tripartite graph $K_{rs,s,s}$ is
$|E|=rss+rss+s^{2}=(2r+1)s^{2}$. From Ref. [26], the genus of the
complete tripartite graph $K_{rs,s,s}$ is given as
\\
\begin{equation}
g=\frac{(rs-2)(s-1)}{2}
\end{equation}
\\
\textbf{Theorem 5.4.1.\hspace*{1mm}}The complete multipartite
graph $K_{m(n)}$, $m\geq 2$, has both an orientable and a
nonorientable self-dual embeddings except in the following cases:
\\
\\
\hspace*{3mm}(1)\hspace*{1mm}When the surface is orientable: $n$
is odd and $m\equiv2$ or $3$ (mod $4$), or when \hspace*{0.9cm}the
graph is $K_{6,6}$, or possibly when the graph is $K_{4(2)}$ or
$K_{6(2)}$.
\\
\\
\hspace*{3mm}(2)\hspace*{1mm}When the surface is nonorientable:
the graph is $K_{n}$, where $n\leq 5$, or when \hspace*{0.9cm}the
graph is $K_{6(3)}$.
\\
\\
\textbf{Proof.\hspace*{1mm}} The Proof of this theorem is found in
[25].
\\
\\
In the complete tripartite graph $K_{rs,s,s}$, we know that
$|E|=(2r+1)s^{2}$. Since in this self-dual embedding (except the
Case $1$ of the above theorem) on an orientable surface the code
minimum distance is three. Thus, the code parameters are given by:
the code minimum distance $d_{min}=3$; the code length is
$n=|E(G)|=(2r+1)s^{2}$ and $k=2g=(rs-2)(s-1)$. Consequently, the
new class of codes with parameters $[[(2r+1)s^{2},(rs-2)(s-1),3]]$
is constructed.
\\
\\
\section{\hspace*{-0.6cm}{.}\ Conclusion}
We considered presentation of four new classes of binary quantum
codes based on self-dual orientable embeddings of the complete
graphs $K_{4r+1}$ and $K_{4s}$, ($r\geq 1$ and $s\geq 2$), complete
bipartite graph $K_{r,s}$, and complete tripartite graph
$K_{rs,s,s}$ by using current graphs and rotation schemes. These
codes are superior to quantum codes presented in other references.
We point out the classes $[[2r(4r+1),2r(4r-3),3]]$ and
$[[2s(4s-1),2(s-1)(4s-1),3]]$ of quantum codes achieve the best
ratio $\frac{k}{n}$. For new classes $[[rs,\frac{(r-2)(s-2)}{2},4]]$
and $[[(2r+1)s^{2},(rs-2)(s-1),3]]$ of codes, the code rate
$\frac{k}{n}$ approaches to $\frac{1}{2}$.

\end{document}